\documentclass[epj,nopacs,final]{svjour}[]
\usepackage{epsfig}
\usepackage{bm}

\newcommand{\g}{\gamma}

\newcommand{\etap}{\eta^{\prime}}

\newcommand{\jpsi}{J/\psi}

\newcommand{\eff}{\varepsilon}

\newcommand{\pip}{\pi^+}
\newcommand{\pim}{\pi^-}
\newcommand{\pin}{\pi^-}
\newcommand{\pio}{\pi^0}

\newcommand{\pbar}{\bar{p}}

\newcommand{\lamlambar}{\Lambda\bar{\Lambda}}

\newcommand{\ppb}{p\bar{p}}

\newcommand{\ar}{\rightarrow}

\newcommand{\bfg}{\begin{figure}[htpb]}
\newcommand{\efg}{\end{figure}}
\newcommand{\bitm}{\begin{itemize}}
\newcommand{\eitm}{\end{itemize}}
\newcommand{\bnum}{\begin{enumerate}}
\newcommand{\enum}{\end{enumerate}}
\newcommand{\btbl}{\begin{table}[htp]}
\newcommand{\etbl}{\end{table}}
\newcommand{\btbu}{\begin{tabular}[htp]}
\newcommand{\etbu}{\end{tabular}}

\newcommand{\bcl}{\begin{center}}
\newcommand{\ecl}{\end{center}}

\newcommand{\beq}{\begin{equation}}
\newcommand{\eeq}{\end{equation}}
\newcommand{\beqr}{\begin{eqnarray}}
\newcommand{\eeqr}{\end{eqnarray}}

\title{\bf\boldmath Study of
  $\jpsi$ decaying into $\omega\ppb$}

\author{
(BES Collaboration)\\
\\
M.~Ablikim\inst{1} \and J.~Z.~Bai\inst{1} \and Y.~Ban\inst{11} \and
X.~Cai\inst{1} \and H.~F.~Chen\inst{16} \and H.~S.~Chen\inst{1} \and
H.~X.~Chen\inst{1} \and J.~C.~Chen\inst{1} \and Jin~Chen\inst{1}
\and Y.~B.~Chen\inst{1}
% ...
 }

\institute {  Institute of High Energy Physics, Beijing 100049,
People's Republic of China\and China Center for Advanced Science and
Technology (CCAST), Beijing 100080, People's Republic of China\and
Guangxi Normal University, Guilin 541004, People's Republic of
China\and Guangxi University, Nanning 530004, People's Republic of
China\and Henan Normal University, Xinxiang 453002, People's
Republic of China\and Huazhong Normal University, Wuhan 430079,
People's Republic of China\and Hunan University, Changsha 410082,
People's Republic of China\and
Liaoning University, Shenyang 110036, People's Republic of China\and
Nanjing Normal University, Nanjing 210097, People's Republic of
China\and Nankai University, Tianjin 300071, People's Republic of
China\and Peking University, Beijing 100871, People's Republic of
China\and Shandong University, Jinan 250100, People's Republic of
China\and Sichuan University, Chengdu 610064, People's Republic of
China\and Tsinghua University, Beijing 100084, People's Republic of
China\and University of Hawaii, Honolulu, HI 96822, USA\and
University of Science and Technology of China, Hefei 230026,
People's Republic of China\and Wuhan University, Wuhan 430072,
People's Republic of China\and Zhejiang University, Hangzhou 310028,
People's Republic of China} \vspace{0.2cm}

\date{Received: date / Revised version: date}
\abstract{ The decay $J/\psi \to \omega p \bar p$ is studied using a
$5.8 \times 10^7$ $J/\psi$ event sample accumulated with the BES~II
detector at the Beijing electron-positron collider. The decay
branching fraction is measured to be $B(\jpsi\ar\omega\ppb)=(9.8\pm
0.3\pm 1.4)\times 10^{-4}$. No significant enhancement near the
$\ppb$ mass threshold is observed, and an upper limit of
$B(\jpsi\ar\omega X(1860))B( X(1860)\ar\ppb)$$
$$< 1.5 \times 10^{-5}$ is determined at the 95$\%$ confidence level,
where $X(1860)$ designates the near-threshold enhancement seen in
the $\ppb$ mass spectrum in $J/\psi \to \gamma p \bar p$ decays.}

\titlerunning{\bf Study of
  $\jpsi$ decaying into $\omega\ppb$ }
\mail{liubj@mail.ihep.ac.cn}
\begin{document}
\maketitle
\section{Introduction}
\label{introduction}

Decays of the $J/\psi$ meson are regarded as being well suited for
searches for new types of hadrons and for systematic studies of
light hadron spectroscopy. Recently, a number of new structures have been
observed in $J/\psi$ decays. These include strong near-threshold mass
enhancements in the $p\pbar$ invariant mass spectrum from
$\jpsi\rightarrow\gamma p\pbar$ decays~\cite{bes1860}, the $p \bar \Lambda$ and
$K^-\bar \Lambda$ threshold enhancements in the $p \bar \Lambda$ and
$K^-\bar \Lambda$ mass spectra in $J/\psi \rightarrow p K^- \bar \Lambda$
decays~\cite{pkl}, the $\omega\phi$ resonance in the $\omega\phi$ mass spectrum
in the double-OZI suppressed
decay $J/\psi\to\gamma \omega\phi$~\cite{goph}, and a new resonance, the $X(1835)$,
in $J/\psi\to\gamma \pi^+\pi^-\eta'$ decays~\cite{x1835}.

The enhancement $X(1860)$ in $\jpsi\rightarrow\gamma p\pbar$
can be fitted with an $S$- or $P$-wave
Breit-Wigner (BW) resonance function. In the case of the $S$-wave
fit, the mass is $1859^{+3}_{-10}$$^{+5}_{-25}$ MeV/$c^2$ and the
width is smaller than 30 MeV/$c^2$ at the 90$\%$ confidence level
(C.L.). It is of interest to note that a corresponding  mass
threshold enhancement is not observed in either $p \bar p$ cross
section measurements or in $B$-meson decays~\cite{B-ppbar}.

This surprising experimental observation has stimulated a number of
theoretical interpretations.
Some have suggested that it is a $p \bar p$ bound state ({\it
baryonium})~\cite{ppbar,theory,gao,yan,baryonium}. Others suggest
that the enhancement is primarily due to final state interactions
(FSI) between the proton and antiproton~\cite{fsi1,fsi2}.

The CLEO Collaboration published results on the radiative decay of
the $\Upsilon (1S)$ to the $p\bar p$ system~\cite{cleoc}, where no
$p \bar p$ threshold enhancement is observed and the upper limit of
the branching fraction is set at $B(\Upsilon(1S)\rightarrow\gamma
X(1860))B(X(1860)\rightarrow p\overline{p})<5\times10^{-7}$ at
90$\%$ C.L.. This enhancement is not observed in BES2 $\psi(2S) \to
\gamma p \bar p$ data either \cite{bes2psip} and the upper limit is
set at $B(\psi(2S)\rightarrow\gamma X(1860))B(X(1860)\rightarrow
p\overline{p})<5.4\times10^{-6}$ at 90$\%$ C.L..

The investigation of the near-threshold
$\ppb$ invariant mass spectrum
in other $J/\psi$ decay modes will be helpful in
understanding the nature of the observed new structures and in clarifying the role of
$\ppb$ FSI effects. If the
enhancement seen in $\jpsi\ar\g\ppb$ is from FSI, it should also be
observed in other decays, such as $\jpsi\ar\omega\ppb$, which  motivated
our study of this channel.  In this paper,
we present results from an analysis of $\jpsi\ar \pip\pin\pio \ppb$
using a sample of $5.8 \times 10^7 J/\psi$ decays
recorded by the BESII detector at the Beijing Electron-Positron
Collider (BEPC).

BES is a conventional solenoidal magnetic detector that is described
in detail in Ref.~\cite{bes}. BESII is the upgraded version of the
BES detector~\cite{besii}. A twelve-layer Vertex Chamber (VC)
surrounds a beryllium beam pipe and provides track and trigger
information. A forty-layer main drift chamber (MDC) located just
outside the VC provides measurements of charged particle
trajectories over $85\%$ of the total solid angle; it also provides
ionization energy loss ($dE/dx$) measurements that are used for
particle identification (PID).  A momentum resolution of $\sigma
_p/p =1.78\%\sqrt{1+p^2}$ ($p$ in GeV/$c$) and a $dE/dx$ resolution
of $\sim$8\% are obtained.  An array of 48 scintillation counters
surrounding the MDC measures the time of flight (TOF) of charged
particles with a resolution of about 200 ps for hadrons. Outside of
the TOF counters is a 12 radiation length, lead-gas barrel shower
counter (BSC), that operates in self quenching streamer mode and
measures the energies and positions of electrons and photons over
$80\%$ of the total solid angle with resolutions of
$\sigma_{E}/E=0.21/\sqrt{E}$ ($E$ in GeV/$c^{2}$),
$\sigma_{\phi}=7.9$ mrad, and $\sigma_{z}=2.3$ cm. External to a
solenoidal coil, which provides a 0.4 T magnetic field over the
tracking volume, is an iron flux return that is instrumented with
three double-layer muon counters that identify muons with momentum
greater than 500 MeV$/c$.

Monte-Carlo simulation is used to determine the mass resolution and
detection efficiency, as well as to estimate the contributions from
background processes. In this analysis, a GEANT3-based Monte-Carlo
program (SIMBES), with a detailed simulation of the detector
performance, is used.  As described in detail in Ref.~\cite{SIMBES},
the consistency between data and Monte-Carlo has been validated
using many physics channels from both $J/\psi$ and $\psi(2S)$
decays.

\section{Analysis of $\jpsi\ar \omega\ppb$, $\omega \to \pip\pin\pio$}
 \label{analysis}
  For candidate
 $\jpsi\ar\pip\pin\pio \ppb$ events, we  require
 four well reconstructed charged
tracks with net charge zero in the MDC and at least two isolated photons in
the BSC.
 Each charged track is required to be well fitted
to a helix, be within the polar angle region $|\cos\theta| <
0.8$,  have a transverse momentum larger than 70 MeV/$c$, and have
a point of closest approach of the track to the beam axis
that is within 2~cm of the
beam axis and within 20~cm from the center of the
interaction region along the beam line.
 For each track, the TOF and $dE/dx$
information is combined to form a particle
identification confidence
level for the $\pi, K$ and $p$ hypotheses; the particle type with
the highest confidence level is assigned to each track. The four
charged tracks are required to consist of an unambiguously identified
$p$, $\bar{p}$, $\pip$ and $\pim$ combination.
 An isolated neutral
cluster is considered as a photon candidate when the angle between
the nearest charged track and the cluster is greater than
5$^{\circ}$, the angle between the $\bar{p}$ track and the cluster
is greater than 25$^{\circ}$~\cite{angpb}, the first hit is in the
beginning of six radiation lengths of the BSC, the difference
between the angle of the cluster development direction in the BSC
and the photon emission direction is less than 30$^{\circ}$, and the
energy deposited in the shower counter is greater than 50~MeV.
 A four-constraint kinematic fit is performed to the hypothesis $J/\psi \to
\ppb\pip\pin\g\g$, and, in the cases where the number of photon
candidates exceeds two, the combination with the  smallest
$\chi^{2}_{\ppb\pip\pin\g\g}$ value is selected. We further require
that $\chi^{2}_{\ppb\pip\pin\g\g} < 20$.

\begin{figure}[htbp]
\centerline{
\hbox{\psfig{file=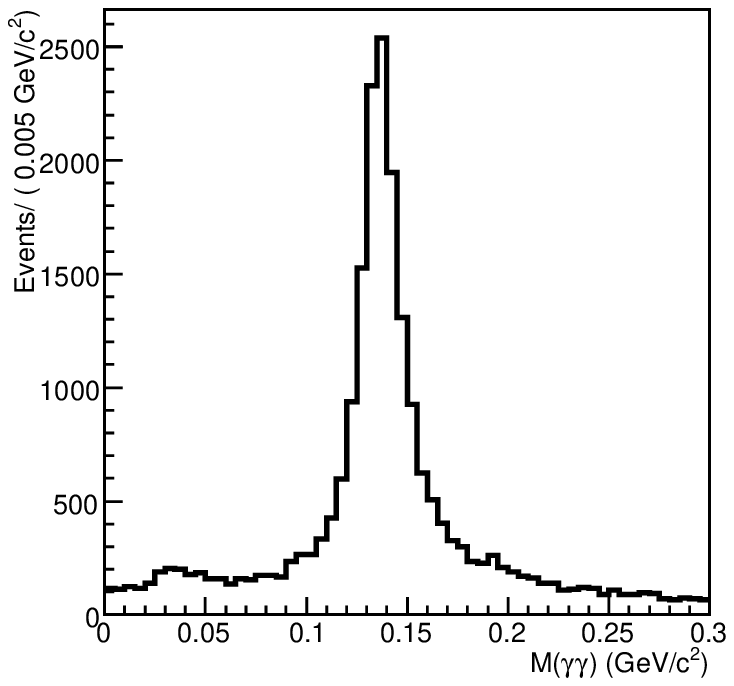,width=7.0cm,height=7.0cm}}}
\caption{The $M_{\g\g}$ distribution for $\jpsi\ar\g\g\pip\pin\ppb$
candidate events.} \label{ompi0}
\end{figure}

\begin{figure}[htbp]
\centerline{
\hbox{\psfig{file=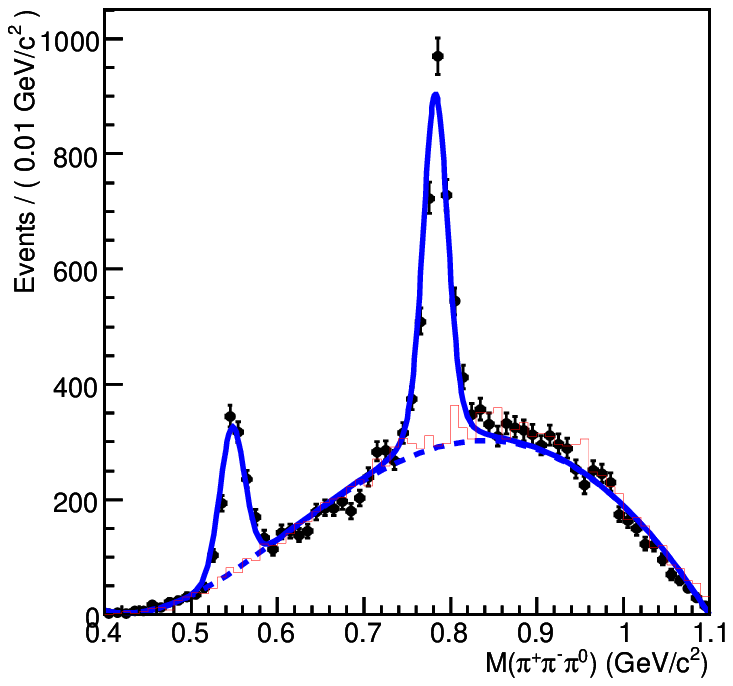,width=7.0cm,height=7.0cm}}}
\caption{The $M_{\pip\pin\pio}$ distribution for
  $\jpsi\ar\pip\pin\pio \ppb$ candidate events. The dots with error bars
are data. The solid histogram is the background estimated from
Monte-Carlo simulation, normalized according to the PDG branching
fractions. The solid curve is the result of a fit described
in the text. The dashed curve is the background polynomial.}
\label{momegafit}
\end{figure}

Figure~\ref{ompi0} shows the $\gamma \gamma$ invariant mass of the
events which survive the above-listed  criteria, where a
distinct $\pio \to \g \g $ signal is evident.
Candidate $\pio$ mesons are selected by requiring
$|M_{\g\g}-m_{\pio}|<0.04$ GeV/$c^2$. After this selection, a total
of 15260 events is retained. The ${\pip\pin\pio}$
invariant mass spectrum for these events is shown
as data points with error bars in Fig.~\ref{momegafit},
where prominent $\omega$ and $\eta$ signals are observed.

The backgrounds in the selected event sample are studied with
Monte-Carlo simulations. We generated $\jpsi
\rightarrow$ $\ppb\pip\pin\pio$ decays as well as a
variety of processes that are  potential sources of background:
$\jpsi \to \ppb\etap (\etap\to\pip\pin\eta)$;
$\ppb\etap (\etap\to\rho^{0}\g)$;
$\ppb\pip\pin$; $\lamlambar\pio$; $\Sigma^0\bar{\Sigma}^0$;
$\Sigma(1385)^-\bar{\Sigma}^+$;
$\eta_c\g$;
$\Delta^{++}\Delta^{--}$;
$\g\ppb\pip\pin$;
$\Delta^{++}\bar{p}\pin$;
$\Lambda\bar{\Sigma}^-\pi^+$ (+ {\it c.c.});
$\Sigma^0\pi^0\bar{\Lambda}$;
$\Sigma(1385)^0\bar{\Sigma}^0$;
$\Delta^{++}\Delta^{--}\pio$; and
$\Xi^0\bar{\Xi}^0$,
in proportion to the branching fractions listed in the
Particle Data Group (PDG) Tables~\cite{pdg2004}.
The main background sources are found to be the decays
$J/\psi \to \Lambda\bar{\Sigma}^-\pi^+$ (+ {\it c.c.}) and
$\Delta^{++}\Delta^{--}\pi^0$. The $\pip\pin\pio$ invariant mass
spectrum for background events that survive the selection criteria
is shown as a solid histogram in Fig.~\ref{momegafit};  here
no signal for $\omega\ppb$  is evident.




The branching fraction for $J/\psi \to \omega p \bar p$ is computed
using the relation
$$B(\jpsi\ar \omega\ppb) =  \frac{N_{obs}}{N_{\jpsi}\cdot \eff\cdot
B(\omega\ar \pip\pin\pio)\cdot B(\pio\ar \g\g)}.$$

Here, $N_{obs}$ is the number of observed events; $N_{\jpsi}$ is the
number of
$\jpsi$ events, $(57.7\pm 2.6)\times 10^6$~\cite{jpsinum};
$\eff$ is the Monte-Carlo determined detection efficiency; and
$B(\omega\ar\pip\pin\pio)$ and $B(\pio\ar\g\g)$ are the
$\omega\ar\pip\pin\pio$ and $\pio\ar\g\g$ branching fractions.

The ${\pip\pin\pio}$ invariant mass spectrum shown in
Fig.~\ref{momegafit} is fitted using an unbinned maximum likelihood
fit with resolution broadened BW functions to represent the $\omega$
and $\eta$ signal peaks. The mass resolutions are obtained from
Monte-Carlo simulation to be 12~MeV$/c^{2}$ for the $\omega$ and and
14~MeV$/c^{2}$ for the $\eta$. The masses and widths of the $\omega$
and $\eta$ are fixed at their PDG values~\cite{pdg2004}. A 4th-order
Chebychev polynomial is used to describe the background. The fit
gives an $\omega$ signal yield of 2449$\pm$69~events. The detection
efficiency from a uniform-phase-space Monte-Carlo simulation of
$\jpsi\ar \omega\ppb$ ($\omega\ar\pip\pin\pio$, $\pio\ar\g\g$) is
$4.9 \pm 0.1)$\%. The branching fraction is determined to be:
$$B(\jpsi\ar\omega\ppb)=(9.8\pm0.3)\times 10^{-4},$$
where the error is statistical only.

\begin{figure}[htbp]
\centerline{
\hbox{\psfig{file=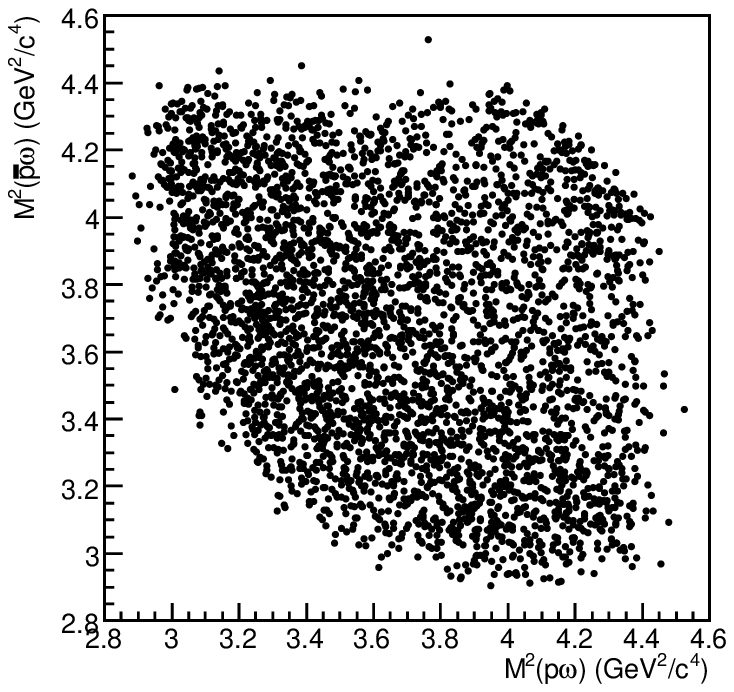,width=7.0cm,height=7.0cm}}}
\caption{The Dalitz plot for $\jpsi\ar\omega\ppb$ candidate events.}
\label{dalitz}
\end{figure}
\begin{figure}[htbp]
\centerline{
\hbox{\psfig{file=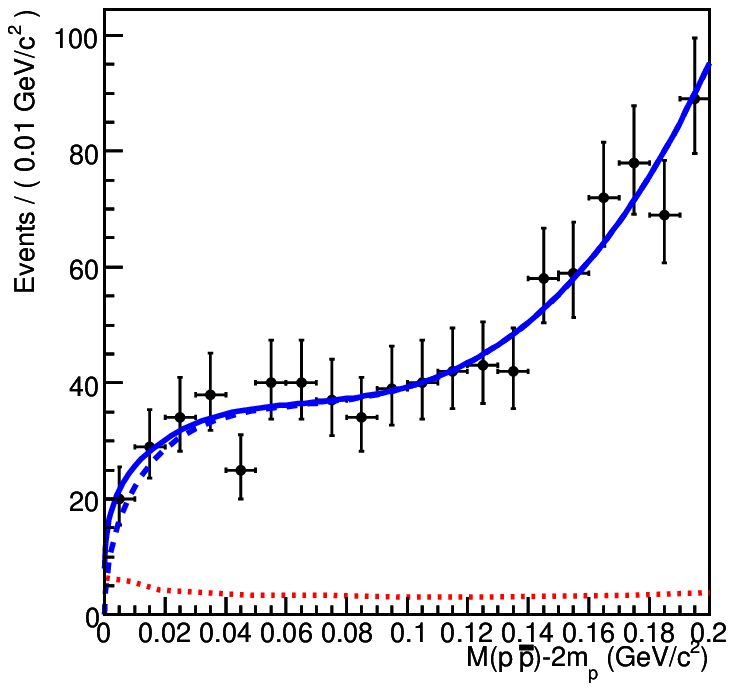,width=7.0cm,height=7.0cm}}}
\caption{The $M_{\ppb}-2m_{p}$ distribution for $\jpsi\ar\omega\ppb$
candidate events. The dots with error bars are data. The solid curve
is the result of fit described in the text. The dashed curve is the
function used to represent the background plus non-resonant
$\omega\ppb$ events. The dotted curve indicates how the acceptance
varies with $\ppb$ invariant mass.} \label{mppbfit}
\end{figure}

We use this sample with $|M_{\pip\pin\pio}-0.783|<0.03$ GeV/$c^2$ to
study the near-threshold region of the $\ppb$ invariant mass
spectrum. Figure~\ref{dalitz} shows a Dalitz plot for the selected
$\jpsi\ar\omega\ppb$ candidates, where no obvious structure is
observed although it is not a uniform distribution.
Figure~\ref{mppbfit} shows
the threshold behavior of the $p \bar p$ invariant mass
distribution.
The dotted curve in the figure indicates how the acceptance varies
with invariant mass.

The backgrounds in the $p \bar p$ threshold region mainly come from
the decays of $J/\psi \to \Lambda\bar{\Sigma}^-\pi^+$ (+ {\it c.c.})
and $\Delta^{++}\Delta^{--}\pi^0$. The $M(\ppb)$ dependence
of this background can be modeled
by appropriately scaled data from the $\omega$
sidebands (0.663 GeV/c$^2$ $<M_{{\pip\pin\pio}}<$0.723 GeV/c$^2$ and
0.843 GeV/c$^2$ $<M_{{\pip\pin\pio}}<$0.903 GeV/c$^2$).

The contributions of sideband and non-resonant $\omega\ppb$ events
can be well described by a function of the form
$$f(\delta) =
N(\delta^{\frac{1}{2}}+a_1\delta^{\frac{3}{2}}+a_2\delta^{\frac{5}{2}})$$
with $\delta\equiv M_{\ppb} - 2m_p $.


In Fig.~\ref{mppbfit},
no significant excess over the background plus non-resonant terms is
evident.
 A Bayesian
approach~\cite{pdg2004} is employed to extract the upper limit on
the branching fraction of $\jpsi\ar\omega X(1860)$. An
acceptance-weighted $S$-wave BW function
$$BW(M) \propto \frac{q^{(2l+1)}k^3}{(M^2-M_0^2)^2-M_0^2\Gamma^2}\cdot\varepsilon(M)$$
is used to represent the low-mass enhancement. Here, $\Gamma$ is a
constant width, $q$ is the momentum of proton in the $\ppb$ rest
frame, $l$ is the relative orbital angular momentum of $p$ and $\bar
p$,  $k$ is the momentum of $\omega$, and $\varepsilon(M)$ is the
detection efficiency obtained from Monte-Carlo simulation.
The mass and width of the BW signal function are fixed to 1860
MeV/c$^2$ and 30 MeV/c$^2$, respectively.
The contributions of background and non-resonant $\omega\ppb$ events
are presented by the function form $f(\delta)$, where the parameters
$a_1$ and $a_2$ are allowed to float.
As shown in
Fig.~\ref{mppbfit}, the solid curve is the fit of the $M_{\ppb}$ -
2$m_p$ with the BW signal function and $f(\delta)$ function
described above.
Using the Bayesian method, the 95\%~C.L. upper limit on the number
of observed signal events is 29. Since the $J^{PC}$ of X(1860) is
unknown, we use simulated events  distributed uniformly in phase
space to determine a detection efficiency of $\jpsi\ar\omega
X(1860)$ ($X(1860)\ar\ppb$, $\omega\ar \pip\pin\pio$, $\pio\ar\g\g$)
of $(4.7 \pm 0.1)$\%.
The upper limit of the branching fraction,
without considering the systematic errors, is then:
$$B(\jpsi\ar\omega X(1860))\cdot B( X(1860)\ar\ppb))$$
$$< \frac{N_{obs}^{UL}}{N_{\jpsi}\cdot \eff\cdot B(\omega\ar
\pip\pin\pio)\cdot B(\pio\ar \g\g)} = 1.2\times 10^{-5}.$$


\section{Systematic errors}
\label{syserrs}

The systematic errors on the branching fractions are mainly due to
uncertainties in the MDC tracking, kinematic fitting, particle
identification (PID), photon detection, background estimation,
the model used to describe hadronic
interactions in the material of the
detector, and the uncertainty of the
total number of $J/\psi$ decays in the data sample.

The systematic error
associated with the tracking efficiency has been carefully
studied~\cite{SIMBES}. The difference of the tracking efficiencies
between data and Monte-Carlo is 2\% per charged track;
an 8\% contribution to the systematic error associated with the
efficiency for detecting
the four-track final state is assigned.
In Ref.~\cite{SIMBES,pnpi}, the efficiencies
for charged particle identification and photon detection are
analyzed in detail. The systematic errors from PID and photon
detection are 2\% per proton (antiproton), 1\% per
pion and 2\% per photon. In this analysis, with four
charged tracks and two isolated photons;
6\% is taken as the systematic
error due to PID and 4\% due to photon detection. The uncertainty due to
kinematic fitting is studied using a number of exclusive
$J/\psi$ and $\psi(2S)$ decay channels that are
cleanly isolated without a kinematic fit~\cite{rhopi,etanpi}. It
is found that the Monte-Carlo simulates the kinematic fit efficiency at
the 5\% or less level of uncertainty for almost all channels tested.
Therefore, we take 5\% as the systematic error due to the kinematic
fit.

The background uncertainties come from the uncertainty of the
background shape. For the branching fraction measurement of
$\jpsi\ar\omega\ppb$, changing the order of the polynomial
background causes an uncertainty in the number of background events.
For the upper limit determination of $\jpsi\ar\omega X(1860)$, the
uncertainty of background shape can be determined by the fitting
results with the background shape fixed to the function form
$f(\delta)$, derived from fitting the scaled $\omega$ sideband data
plus phase-space generated $\omega\ppb$ MC events. Respectively, 5\%
and 10\% are taken as the systematic errors due to the background
uncertainties in the branching fraction measurement of
$\jpsi\ar\omega\ppb$ and the upper limit determination of
$\jpsi\ar\omega X(1860)$.

Different simulation models for the hadronic interactions in the
material of the detector (GCALOR/FLUKA)~\cite{gcalor,fluka} give
different efficiencies. Respectively, 4.8\% and 11.4\% are taken as
the systemic errors due to the different hadronic models in the
branching fraction measurement of $\jpsi\ar\omega\ppb$ and the upper
limit determination of $\jpsi\ar\omega X(1860)$. In addition, if the
$J^{P}$ of X(1860) is $0^{-}$ , the angular distribution of the
$\omega$ would be $1+cos^{ 2}\theta$. A Monte-Carlo sample generated
with the $\omega$ produced with a $1+cos^{2}\theta$ distribution and
 a uniform distribution for the X(1860) decay into $\ppb$
results in an 8.5\% reduction in detection efficiency.  This
difference is taken as the systematic error associated with the
production model.

The branching fractions of $\omega\ar \pip\pin\pio$ and
$\pio\ar\g\g$ are taken from the PDG tables. The errors of the
intermediate
decay branching fractions, as well as the uncertainty of the number of
$J/\psi$ events~\cite{jpsinum} also result in the systematic
errors in the measurements.

The systematic errors from the different sources are listed in Table
\ref{syserr}. The total systematic errors for the branching
fractions are obtained by adding up all the systematic sources in
quadrature.

\begin{table}[htpb]
\caption{Systematic error sources and contributions (\%).}
\begin{center}
\begin{tabular}{ |l|c|c|c}
\hline &B($\jpsi\ar\omega\ppb$)
&Upper Limit \\
\hline
Tracking efficiency & 8   & 8 \\

Photon efficiency & 4   & 4\\
Particle ID & 6   & 6 \\
Kinematic fit &5  &5 \\
Background uncertainty &5  &10  \\
Hadronic model & 4.8  & 11.4  \\
Production model & -  & 8.5  \\
Intermediate decays & 0.8  & 0.8 \\
Total $\jpsi$ events& 4.7    & 4.7\\
\hline
Total systematic error & 14.6  &21.6 \\
\hline
\end{tabular}
\label{syserr}
\end{center}
\end{table}

\section{Summary}
With a $5.8 \times 10^7 \jpsi$ event sample in the BESII
detector, the branching
fraction  $\jpsi\ar\omega\ppb$ is measured as:
$$B(\jpsi\ar\omega\ppb)=(9.8\pm 0.3\pm 1.4)\times 10^{-4}.$$

No obvious near-threshold $\ppb$ mass enhancement in
$\jpsi\ar\omega\ppb$ is observed, and the FSI interpretation of the
$p \bar p$ enhancement in $J/\psi \to \gamma p \bar p$ is
disfavored. A conservative estimate of the upper limit is determined
by
lowering the efficiency by one standard deviation.
In this way,  a 95\% confidence level upper limit on the branching
fraction
$$B(\jpsi\ar\omega X(1860))\cdot B( X(1860)\ar\ppb))< 1.5 \times 10^{-5}$$
is determined. The absence of the enhancement $X(1860)$
in $J/\psi \to \omega p \bar p$, $\Upsilon(1S) \to \gamma p \bar p$ and
$\psi(2S) \to \gamma p \bar p$ also indicates its similar production
property to that of $\eta'$ \cite{ichep06,klempt}, {\it i.e.}, $X(1860)$
is only largely produced in $J/\psi$ radiative decays.


\section{Acknowledgments}

The BES collaboration thanks the staff of BEPC and computing
center for their hard efforts. This work is supported in part by
the National Natural Science Foundation of China under contracts
Nos. 10491300, 10225524, 10225525, 10425523, 10625524, 10521003,
the Chinese Academy of Sciences under contract No. KJ 95T-03, the
100 Talents Program of CAS under Contract Nos. U-11, U-24, U-25,
and the Knowledge Innovation Project of CAS under Contract Nos.
U-602, U-34 (IHEP), the National Natural Science Foundation of
China under Contract No. 10225522 (Tsinghua University), and the
Department of Energy under Contract No. DE-FG02-04ER41291 (U.
Hawaii).



\begin{thebibliography}{xx}
\bibitem{angpb} The annihilation of $\pbar$ would produce many photons. Making the
requestion of the angles between $\pbar$ and photons more strict can
effectively reduce the background from $\pbar$ annihilation.
\end{thebibliography}
\end{document}